\documentclass{appolb}
\usepackage{graphicx}

\begin{document}
\title{Improved $K_S$ tagging procedure and its impact on physics at KLOE-2%
}
\author{M. Silarski on behalf of the KLOE-2 Collaboration
\address{Institute of Physics, Jagiellonian University, PL-30-059 Cracow, Poland}
\address{Laboratori Nazionali di Frascati dell'INFN, Frascati, Italy}
}
\maketitle
\begin{abstract}
The KLOE experiment at the DA$\Phi$NE $\phi$-factory performed precise studies of charged and
neutral kaon physics, low energy QCD, as well as tests of CP and CPT invariance.
For the new run the KLOE has been upgraded by adding new
tagger systems for the $\gamma\gamma$ physics, the inner tracking chamber
and two calorimeters in the final focusing region.
We are also improving on kaon identification techniques, in particular algorithms
for the $K_S$ meson tagging. In this article we discuss the impact of the improved
tagging procedure on studies of the $K_S$ decays.
\end{abstract}
\section{Introduction}
The $\phi$ meson produced in $e^+e^-$ collisions at DA$\Phi$NE is in a pure $J^{PC} = 1^{- -}$ state.
Thus, neutral kaon pairs are in an antisymmetric state which can be expressed
in the $\phi$ rest frame as:
\begin{equation}
\left | i \right \rangle =N \cdot \left[\left|K_S(\vec{p})\right\rangle\left|
K_L(-\vec{p})\right\rangle -\left|K_L(\vec{p})\right\rangle\left|K_S(-\vec{p})\right\rangle\right]~,
\label{ksklstate}
\end{equation}
where $\vec{p}$ denotes the momentum of each kaon and $N$ is a normalization factor~\cite{DiDomenico}.
Since $e^+e^-$ beams collide in the horizontal plane at small angle, $K_S$ and $K_L$ are produced
almost back-to-back, with total momentum, $P_T \sim 15$~MeV/c.
Therefore, observation of a $K_L$ ($K_S$) meson ensures (tags) the presence of the $K_S$ ($K_L$)
flying in the opposite direction and the kinematical closure can be used to determine the momentum
of tagged kaons. 
Thus, at DA$\Phi$NE we obtain pure $K_S$ and $K_L$ beams with
precisely known momenta and flux, which can be used to measure absolute branching fractions~\cite{kloe2008}.
The tagging is performed mainly by the reconstruction of the $K_L$
interaction in the calorimeter (``$K_L$ crash''), which provides a very clean
identification of the $\phi \to K_S K_L$ events. The other method is based on reconstruction
of the $K_L$ decay inside the drift chamber which may significantly increase the tagging efficiency. 
\section{$K_S$ tagging via detection of the $K_L$ in the KLOE calorimeter}
\begin{figure}
\centerline{%
\includegraphics[width=6.5cm]{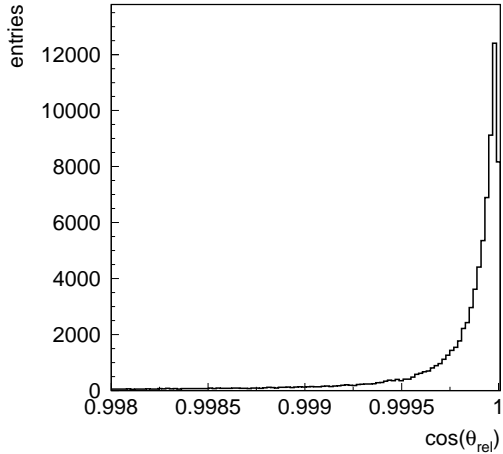}}
\caption{Distribution of the angle between true and reconstructed $K_L$ direction
for a sample of simulated $\phi \to K_L K_S$ events. The full width at half maximum of the
distribution corresponds the the accuracy of $\sim$~1$^\circ$.}
\label{Fig1a}
\end{figure}

\begin{figure}
\centerline{%
\includegraphics[width=6.5cm]{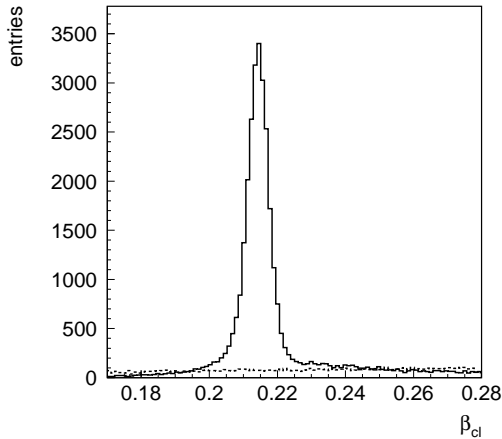}}
\caption{Simulated distributions of the reconstructed $K_L$ velocity $\beta_{cl}$
for a sample of the $\phi \to K_S K_L$ events (solid histogram) and background
(dashed histogram).}
\label{fig1}
\end{figure}

At KLOE about 60$\%$ of produced $K_L$ mesons reach the calorimeter
where they can interact~\cite{kloe2008}.
Thanks to the excellent time resolution of the KLOE calorimeter
and the low velocity of kaons one can use the Time of Flight
technique to tag the $K_S$ meson. Adding the information about the position of the energy release
($K_L$ cluster), the direction of the $K_L$ flight path can be determined with a good precision.
In Fig.~\ref{Fig1a} we present distribution of cos$\theta_{rel}$, i.e., cosine of an angle
between reconstructed and true $K_L$ direction for a sample of simulated $\phi \to K_S K_L$ events.
The full width at half maximum corresponds to about $1^\circ$.
This allows to determine $K_L$ kinematics and, knowing the the total energy and momentum from
the analysis of Bhabha scattering events, to determine the four-momentum of the tagged $K_S$ meson.\\ 
The identification of the $K_L$ interaction in the calorimeter is performed after
tracks reconstruction and after applying the track-to-cluster association procedure.
A sequence of cuts is then
applied to reject events with $K_L$ decay inside the drift chamber~\cite{km146}. 
For each event we look for the $K_L$ clusters in the calorimeter
taking into account only clusters not associated to any track. For these clusters we calculate
the particle velocity defined in the laboratory frame as
$\beta_{cl} = R_{cl}/(c\cdot t_{cl})$, where $R_{cl}$ is the distance from the $e^+e^-$
interaction point to the reconstructed position
of the cluster center, $t_{cl}$ stands for the measured time of flight of the particle and $c$ is
the speed of light. It is used to select clusters corresponding to $K_L$ with
$\beta_{cl} \sim 0.22$ (see Fig.~\ref{fig1}).
To reject delayed clusters due to charged pions for which the track-to-cluster association
procedure failed, we require an energy deposition of at least 100~MeV.
Kaons from the $\phi$ decay are mostly emitted at a large polar angle
so that the background can be additionally suppressed selecting only ``$K_L$ crash'' clusters
in the barrel calorimeter~\cite{Flavio}.
The small remaining background contamination originates from $\phi \to K^+K^-$ decays and
cosmic muons entering KLOE through the intersection between the barrel and endcap calorimeters.
As one can see in
Fig.~\ref{fig1} this contamination is characterized by a flat $\beta_{cl}$ distribution.\\
The efficiency of this tagging procedure depends on the requirements on $K_L$ velocity
and cluster energy, and is in the range of
23-34\%~\cite{silar,Ambrosino:2008zi,newlim,Ambrosino:2007ab,Ambrosino:2006si,Ambrosino:2005iw}.
It is worth mentioning that, according to the KLOE Monte Carlo simulations, about 30\% of
$K_L$ interactions reconstructed in the calorimeter fulfill the KLOE trigger conditions
allowing for a search of the $K_S \to invisible$ decays which, if observed, would be
an unambiguous signal of physics beyond the Standard Model~\cite{Gninenko}.
\section{$K_S$ tagging with $K_L$ charged decay reconstruction}
Charged $K_L$ decays in the KLOE drift chamber which are rejected by the
``$K_L$ crash'' algorithm can also be used to tag $K_S$~\cite{km160}. This can be done by
looking for an isolated vertex or chain of vertices
inside the drift chamber which are reconstructed outside a sphere of 30~cm radius around the
interaction point. In addition each track associated to the vertex should not point to the
interaction region. More detailed description of the analysis cuts can be found in Ref.~\cite{km160}.\\
Efficiency of this tagging procedure was studied with Monte Carlo simulations for the main $K_S$ decay
channels. About 30\% of generated $K_L$ mesons decay inside the KLOE drift chamber and about
55-60\% of these decays fulfill the tagging conditions with a small dependence on the $K_S$
decay channel (tag bias).
Main background source for this tagging algorithm originates from the $\phi \to K^+K^-$ and
$\phi \to \pi^+\pi^-\pi^0$ decays giving a few percent contamination.  
\section{Conclusions and outlook}
$K_S$ tagging with $K_L$ charged decay reconstruction applied together with the $K_L$-crash
algorithm can increase the statistics for $K_S$ branching ratio measurements by a factor of 1.5,
which is a significant improvement in view of rare $K_S$ decays studies. However, further
studies to reject residual contamination from $\phi \to K^+K^-$ are needed for full exploitation
of the additional sample. In particular, the evaluation of the
impact of tagging with $K_L$ charged decay reconstruction on rare $K_S$ decays measurements,
such as the $K_S \to \pi^0\pi^0\pi^0$ and $K_S \to \pi^+\pi^-\pi^0$ decays is a first objective
of these studies.
Improved kaon identification techniques are also important in view of data which have been collected
with the KLOE-2 apparatus equipped with the inner tracker~\cite{it}, new scintillation calorimeters~\cite{det4,det5}
and tagging detectors for $\gamma\gamma$ physics~\cite{bab,Arch}.
These measurements will allow to refine and extend the KLOE program on kaon physics and tests of
fundamental symmetries as well as the quantum interferometry~\cite{camelia}. 
\section*{Acknowledgements}
The author is grateful to Caterina Bloise for her valuable comments and corrections of the manuscript.\\
This work was supported in part by the EU Integrated Infrastructure Initiative Hadron Physics Project under
contract number RII3-CT- 2004-506078; by the European Commission under the 7th Framework Programme through
the `Research Infrastructures' action of the `Capacities' Programme, Call: FP7-INFRASTRUCTURES-2008-1,
Grant Agreement No. 227431; by the Polish National Science Centre through the Grants No.  
DEC-2011/03/N/\\ST2/02641, 
2011/01/D/ST2/00748,
2011/03/N/ST2/02652,
2013/08/M/\\ST2/00323,
2013/11/B/ST2/04245,
and by the Foundation for Polish Science through the MPD programme and the project HOMING PLUS BIS\\/2011-4/3.

\end{document}